# Electrostatic Photoluminescence Tuning in All-Solid-State Perovskite Transistors.

(March 17, 2026)

Vladimir Bruevich[1,§], Dmitry Maslennikov[2,§], Beier Hu[2], Artem A. Bakulin[2], Vitaly Podzorov[1,*]

[1] Department of Physics & Astronomy, Rutgers, the State University of New Jersey, NJ, USA.

[2] Department of Chemistry and Centre for Processible Electronics, Imperial College London, UK.

[*] Email: podzorov@physics.rutgers.edu

[§] These authors contributed equally to this work.

We demonstrate an all-solid-state semiconductor device, based on epitaxial single-crystalline metal-halide perovskites, enabling reversible control of a perovskite's photoluminescence with a gate voltage. Fundamentally distinct from electroluminescent diodes, such a *photoluminescence field-effect transistor* uses the gate electric field to electrostatically modulate the interfacial density of mobile charges, thereby affecting the radiative and non-radiative recombination channels of photocarriers. Varying the gate voltage in such transistors efficiently changes the rate of non-radiative interfacial recombination and modulates the photoluminescence intensity by 65 – 98 % (depending on temperature). At favorable gating, nearly complete elimination of non-radiative losses can be achieved. This functionality, coupled with the strong visible-range absorption and emission, possible due to the high absorption coefficient, as well as controllable thickness and macroscopically homogeneous morphology of epitaxial perovskite films, leads to high *external* photoluminescence quantum efficiencies realized in large-area, thin-film devices. Such high-efficiency, scalable, electrostatically tunable optoelectronic switches broaden the potential applications of metal-halide perovskites in photonics and optoelectronics.

**Introduction**

Tuning physical properties of emerging electronic materials with an "electric knob" has been an important goal in materials physics and device engineering.[1] Unlike irreversible chemical modifications, reversibly tuning materials' optoelectronic properties with external stimuli, such as, e.g., uniaxial strain,[2] or magnetic and electric fields,[1, 3, 4] significantly widens the range of potential applications. A notable example is an electrostatic tuning of electric conductivity in field-effect transistors (FETs) via modulation of mobile carrier density at the semiconductor surface, occurring without introducing chemical or structural disorder. Not only have FETs been pivotal to the birth of the information age,[5] they have also enabled the basic research into the charge transport physics of a variety of materials.[1, 6-8]

While controlling the electric conductivity of materials in FETs is a vast area of research, reversibly tuning their optoelectronic properties (e.g., the photoluminescence (PL) or photoconductivity (PC)) in a similar manner is underexplored. In the past, researchers used static electric field in silicon-based metal-oxide-semiconductor (Si MOS) devices to alter the Shockley-Reed-Hall surface recombination rates by redistributing near-surface populations of electrons and holes.[9-14] Although these pioneering works made a conceptual leap forward and provided a tool for the experimental determination of surface recombination velocities and carrier lifetimes, Si is an indirect bandgap semiconductor (with a low bandgap of 1.12 eV in the near infra-red), making it inefficient light emitter.

Layered inorganic semiconductors, such as transition-metal dichalcogenides (TMDs), was another promising material system to implement electrostatic control of emissive and optoelectronic properties. TMDs exhibit efficient in-plane charge carrier transport and naturally dangling-bond-free exfoliated surfaces, crucial for the realization of FETs and other devices.[15-18] Although monolayers of these materials are promising for optoelectronics due to their direct bandgap,[19] they are typically excitonic even at room temperature. The PL of excitonic materials, including organic semiconductors and inorganic monolayer materials,[20, 21] originates from a radiative recombination of singlet excitons and exhibits intrinsic intensity-dependent exciton-exciton or exciton-charge quenching, thus limiting their use to low power densities.[21-23] Additionally, although it was possible to control radiative and non-radiative recombination rates in monolayer FETs,[21] individual layers typically absorb only a small fraction of incident light (a few %),[24, 25] which limits the maximum achievable light tunability possible with such devices. On the contrary, in non-excitonic direct-bandgap semiconductors, the PL stems from a radiative recombination of free electrons and holes, which is mainly limited by non-radiative interaction with defects, making highly-ordered and pure materials of this type more feasible for practical applications.

Metal-halide perovskites (MHPs) have recently emerged as semiconducting materials that combine modest charge transport with excellent optical properties, promising for an array of applications. High PL yield,[26, 27] non-excitonic photoexcited state at room temperature,[28] sufficiently high charge-carrier mobilities,[29-35] apparent defect tolerance,[36] and sharp absorption edge,[27, 37] combined with a remarkable synthetic tunability that enables processable and scalable MPHs of varied dimensionality, make these materials an ideal platform for exploring the effect of electrostatic gating on the photoluminescence. Although a control of PL properties of MHPs in ionic-liquid gated transistors has been demonstrated, possible electrochemical interactions at the perovskite-ionic liquid interface could not be fully ruled out.[38]

To electrostatically control a bulk PL in a FET, the semiconductor's bulk and surface must be of exceptional quality (free from chemical and structural disorder) to allow mobile photocarriers reach the field-induced accumulation layer at the surface and interact with mobile gate-induced carriers. Recently, we reported on epitaxial single-crystalline cesium lead bromide (CsPbBr$_3$) perovskite FETs exhibiting *intrinsic* (i.e., not limited by static disorder) charge carrier transport.[30] Such all-solid-state devices are ideal for exploring an electrostatic PL tuning, because they combine high-quality bulk and interfaces with the desired high PL quantum yield of the material.[39]

In this work, by taking advantage of excellent charge transport and light emission properties of epitaxial single-crystalline CsPbBr$_3$, we demonstrate an efficient and reversible control of the photoluminescence emission from CsPbBr$_3$ transistors with a gate voltage. Such all-solid-state *photoluminescence transistors* (PLTs) enable PL tuning by modulating the mobile charge density at the semiconductor-dielectric interface with a gate voltage. We have developed an analytical model that appears to correctly capture the main observed trends of the device operation. Our results introduce a novel optoelectronic device, where the PL yield of a three-dimensional, non-excitonic, direct-bandgap semiconductor is electrostatically controlled with a gate voltage in a FET geometry. Such a tunable optical switch is promising for a range of applications in optoelectronics, including lasing, displays, telecommunications, optical integrated circuits, and sensing.

**Results**

In this work, we report an electrostatic *photoluminescence gating* in all-solid-state metal-halide perovskite field-effect transistors. A significant and reversible modulation of the PL intensity, $I_{PL}$, is recorded in epitaxial single-crystalline CsPbBr$_3$ transistors with a gate voltage, $V_G$, applied to a semi-transparent gate electrode. **Figure 1a** schematically shows our experimental geometry, where a top-gated CsPbBr$_3$ transistor is imaged *operando* via a PL microscope (for details on device fabrication and measurements, see Methods and Supplementary sec. 1).[1] **Figure 1b** shows a series of PL images of the PLT's channel recorded through the (semi)transparent gate and gate dielectric at four gate voltages, $V_G$ = 50, 30, -10, and -50 V, demonstrating a monotonically increasing brightness of the channel (for the complete set of PL images recorded with a finer step of $\Delta V_G = \pm 10$ V, as well as the corresponding video of the PL evolution, see Supplementary sec. 2 and Mov. S1). Especially in the movies, the effect can be easily seen by a naked eye: decreasing $V_G$ to - 50 V significantly enhances the PL emission in the gated area of the crystal, while driving the transistor into a depletion at $V_G$ = 50 V switches the PL off completely. More detailed analysis of the PL movies (including a stray-light background subtraction) shows that varying $V_G$ in such transistors allows modulating the PL intensity by approximately 97.7% (at - 20 °C). Due to the well-known superior performance of *parylene* conformal coatings as a gate insulator resilient against leaks (see, e.g., Refs.[40, 41]), the observed PL gating effect is purely electrostatic in nature. Indeed, the gate leakage current in CsPbBr$_3$ transistors in this $V_G$ range is negligible,[30] thus ruling out parasitic electroluminescent phenomena. Therefore, the gate electrode controls the PL intensity of the channel in a similar way it controls the source-drain electric current, $I_{DS}$, flowing through the FET channel.

---

[1] For Supplementary materials, please contact the corresponding author at podzorov@physics.rutgers.edu

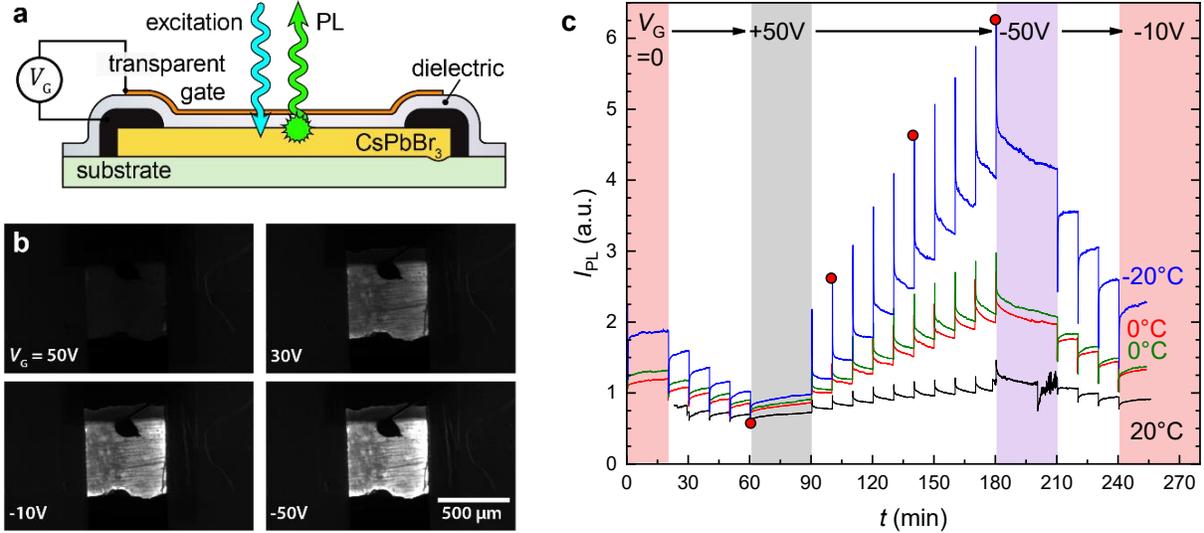

**Figure 1.** *Operando* **PL microscopy of metal-halide perovskite field-effect transistors.** (**a**) A sketch of the experiment showing a top-gated perovskite *photoluminescence transistor* (PLT) with a transparent gate electrode under a PL microscope. The imaging is performed utilizing a *cw* blue (440 - 490 nm) excitation and a 495 nm long-pass emission filter, with a CCD camera for the emission detection (for details, see Methods and Supplementary sec. 1.3). (**b**) Several representative top-view microscopic PL images of a single-crystal CsPbBr$_3$ PLT recorded at $V_G$ = 50, 30, -10, and -50 V (for the complete set of images and a movie, see Supplementary sec. 2 and Mov. S1). The images were recorded at - 20 °C. (**c**) A response of the channel-area averaged PL intensity, $I_{PL}$, of the PLT to $V_G$ very slowly swept in steps of $\Delta V_G = \pm 10$ V applied every ~ 10 min. To verify reversibility, measurements were performed at the following stabilized temperatures: $T$ = 20 °C (black line), 0 °C (red line), -20 °C (blue line), and then back to 0 °C (green line). The red dots correspond to the PL images of the PLT's channel shown in panel **b**.

To better understand the PL gating effect, we have performed PL imaging of an operating CsPbBr$_3$ transistor at different temperatures ($T$ = 20, 0, and -20 °C), with $V_G$ very slowly varied in a stepwise fashion between -50 and 50 V in steps of $\Delta V_G = \pm 10\ V$ applied every 10 min. Digital PL images recorded at a rate of one frame per 500 ms were then analyzed by calculating $I_{PL}$ averaged over the transistor's channel. **Figure 1c** shows the resultant evolution of $I_{\text{PL}}$ with $V_{\text{G}}$, revealing that stepwise changes in the gate voltage initially cause big but transient spikes in the PL intensity. After the spikes decay, the PL equilibrates to a persistent $V_{\text{G}}$-dependent plateau. Cooling the transistor to just -20 °C leads to a noticeable increase in both the absolute PL intensity and the $V_{\text{G}}$-induced change, $\Delta I_{\text{PL}}$ (Fig. 1c).

**Figure 2** presents the effect of a single $V_{\text{G}}$ step recorded at a higher temporal resolution of one frame every 100 ms. The normalized, channel-area averaged PL intensity vs. time, $I_{\text{PL}}(t)/I_{\text{PL}}^{\max}$, exhibits a strong, rapid initial rise at $\Delta t$ = 0 when a step of $\Delta V_{\text{G}} = -10$ V is applied, followed by a slower exponential relaxation lasting for 2 - 7 s, with the signal eventually leveling off to a nearly constant persistent background. The inset in Fig. 2 is a double-log plot of the corresponding $I_{\text{PL}}(\Delta t)$ traces fitted with a single exponential decay and revealing the time constants of $\tau_{\text{PL}}$ = 0.54 and 1.97 s for $T = 0$ and $-20$ °C, respectively. The key observation here is that the time constant of the PL decay noticeably increases with cooling. The very long time constant of the decay (~ seconds), the increase of $\tau_{\text{PL}}$ with cooling, and the fact that a small relative reduction in temperature ($\Delta T/T \approx 7.5$ %) leads to such a significant slowdown of the PL dynamics ($\Delta \tau_{\text{PL}}/\tau_{\text{PL}} \approx 265$ %) around room temperature are all suggestive of a decay associated with an ionic redistribution in the interfacial region of the perovskite lattice occurring in response to a changing gate electric field.

On the contrary, the rapid initial rise of the PL right at the onset of the $V_{\text{G}}$ steps likely corresponds to the fast electronic response involving the interaction of mobile photogenerated carriers with interfacial gate-induced holes in the transistor's accumulation channel. Another observation supporting this assertion is the magnitude, $\Delta I_{\text{PL}}^{\text{spike}}$, of the rapid initial raise in $I_{\text{PL}}$ in response to a $V_G$ step that becomes greater at lower temperatures (compare the size of the spikes in Fig. 1c at different temperatures). This increase could be primarily associated with an increase in the band-like charge carrier mobility, $\mu$, occurring in single-crystalline perovskites upon cooling,[30-35] which leads to a stronger effect of $V_G$ on the radiative recombination. Note that the fast initial rise in PL cannot be directly attributed to the electric-field-driven ion migration in the lattice, because (**a**) such thermally-activated ionic hopping is expected to rapidly diminish with cooling, while we observe an increase in $\Delta I_{PL}^{spike}$ at lower temperatures, and (**b**) under a *cw* photoexcitation, any changes in the gate electric field are expected to be quickly screened by rapidly redistributing populations of photocarriers present in the material as they are much more mobile than ionic species and thus reorganize quickly in response to a $V_G$ step. We therefore hypothesize that the true electronic PL gating effect comprises not just the persistent PL plateau seen in Fig. 2 at longer times but corresponds to the strong initial rise in $I_{PL}$ occurring immediately upon the application of $V_G$ steps. The true rise time of these spikes cannot be currently resolved, because it is shorter than the sampling interval of our CCD camera-based imaging system required to obtain proper spatial resolution; possible dependence of this rise time on temperature, carrier mobility, or excitation intensity could be an interesting subject for follow-up research.

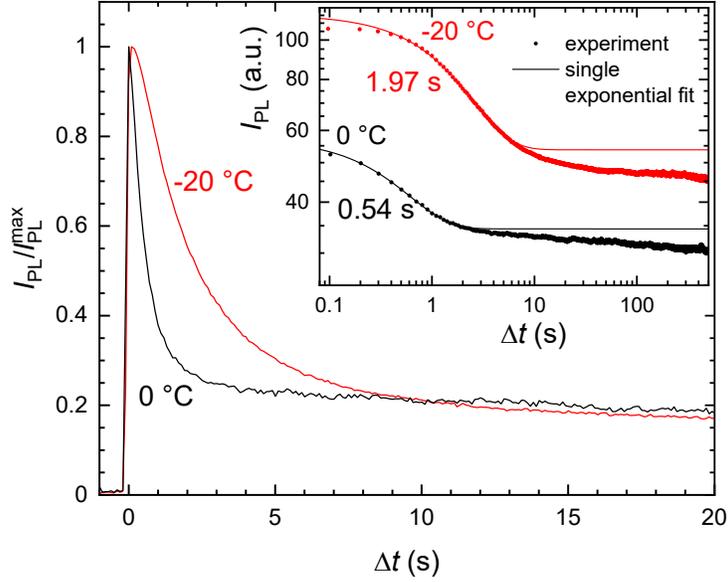

**Figure 2. High-temperature kinetics of the PL gating effect in perovskite photoluminescence transistors.** Main panel (double-linear plot): a channel-area averaged and normalized to the peak PL intensity of a single-crystalline CsPbBr$_3$ PLT recorded as a function of time at a rate of 10 frames/s, while a gate-voltage step of $\Delta V_G = -10$ V was applied at $\Delta t = 0$. Measurements were carried out at two stabilized temperatures near room temperature, $T = 0$ °C and $-20$ °C (black and red lines or symbols, respectively). Inset (double-log plot): the same measurements in a non-normalized form, shown to a longer time (symbols), with the initial fast decay fitted with a single exponential-decay function (solid lines); the decay time constants are indicated.

To test this hypothesis further, we have performed PL gating measurements at even lower temperatures. Cooling should help discriminate electronic processes from ionic drifts, as the former is expected to become faster, while the latter is expected to freeze at sufficiently low $T$. Epitaxial single-crystalline CsPbBr$_3$ transistors are well-suited for such measurements, because they exhibit exceptionally low hysteresis even at room temperature and a negligible hysteresis below ~ 200 K.[30] We have thus performed a series of additional PL gating measurements at $-95\,°C$. **Figure 3a** shows the kinetics of the PL response of a CsPbBr$_3$ PLT to a $V_G$ changing between $-50$ and $50$ V in steps of $\Delta V_G = \pm 10$ V (for details see Methods and Supplementary sec. 1.3).

It is apparent from Fig. 3a that: (a) a noticeable and reversible PL gating effect is present; (b) after each step in $V_G$, the PL changes and stays at a plateau; (c) the relatively fast exponential decay seen at high temperatures is gone (only a very slow, voltage-dependent drift and fluctuations can be seen at the plateaus); and (c) sweeping $V_G$ back and forth reveals some hysteresis in the PL response. The $I_{PL}(V_G)$ dependence, normalized to the relaxed zero-gate PL intensity, is plotted in Fig. 3b. Open symbols correspond to the PL intensity calculated by averaging the experimental data for each $V_G$. The observed hysteresis in $I_{PL}(V_G)$ is likely associated with the effect of a gate-assisted photoinduced charge transfer at the semiconductor-dielectric interface - the phenomenon originally observed in single-crystalline organic FETs and later found to be ubiquitous.[42, 43] As shown in Supplementary sec. 3, increasing the excitation flux density leads to larger hysteresis. In addition, we have also collected the corresponding PL spectra of the operating CsPbBr$_3$ PLT at different $V_G$, revealing that, within our experimental accuracy, changing $V_G$ only affects $I_{PL}$ but not the spectral shape or peak position ($\lambda_{max}$). The changes in $\lambda_{max}$ occurring over the full-range $V_G$ variations are estimated to be not greater than 0.14 nm (Supplementary sec. 4).

The observed modulation of the PL intensity with the gate voltage, $I_{PL}(V_G)$, resembles a field-effect transistor action in electrical measurements. Indeed, in a *p*-type FET, decreasing the gate voltage below the threshold voltage, $V_G < V_T$, leads to an increasing concentration of mobile holes in the channel, thus increasing the channel's conductivity, while applying a sufficiently high gate voltage above the threshold, $V_G > V_T$, depletes the channel of mobile holes, switching the conduction off.[44] According to the electrical measurements of single-crystal CsPbBr$_3$ FETs,[30] these devices indeed operate as classic *p*-type FETs, though with a positive threshold voltage. As previously shown, such a $V_T > 0$ originates from a mild interfacial charge-transfer doping of the perovskite surface caused by parylene-*N*, leading to these FETs being *on* at $V_G = 0$.[30] Nevertheless, besides a solid electrostatic onset shift, these devices operate as textbook FETs. The observed resemblance between the behavior of the PLTs and electrical FETs suggests that the modulation of $I_{PL}$ with $V_G$ is associated with mobile holes accumulated in the transistor's channel. Saturation of the $I_{PL}(V_G)$ dependence at $V_G < 0$ at low temperatures (Fig. 3b) provides additional clues for developing a microscopic understanding of the effect as described below.

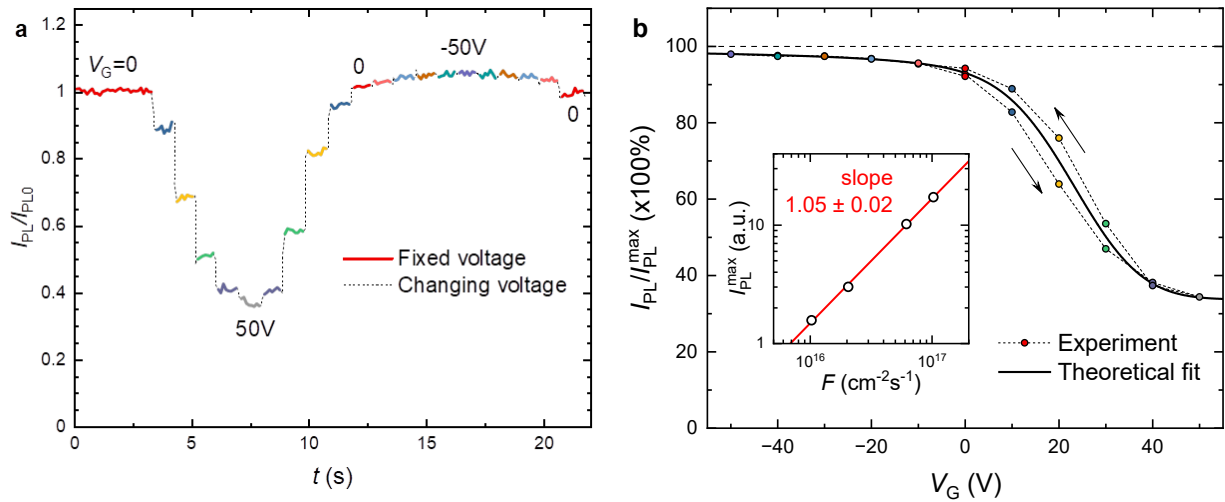

**Figure 3. Modulation of the PL intensity of a perovskite photoluminescence transistor with $V_G$ at -95 °C.** (**a**) A normalized, channel-area averaged PL intensity of a single-crystalline CsPbBr$_3$ transistor measured while $V_G$ is swept in a sequence $0 \rightarrow 50\ V \rightarrow -50\ V \rightarrow 0$ in $\Delta V_G = \pm 10\ V$ increments. Colors represent different $V_G$ levels. (**b**) Main panel: a PL transfer characteristic of a CsPbBr$_3$ transistor, showing the dependence of the normalized PL intensity on the gate voltage (open circles), with the arrows indicating

the direction of hysteresis (counterclockwise). The colored solid circles represent the averages over the hysteresis for each $V_G$. A fit with the theoretical model described in the text (Eqs. 5 and 6) is shown with a black solid line. Inset: the maximum experimental PL intensity, $I_{PL}(V_G = -50\ V)$, plotted as a function of photoexcitation flux density, revealing a linear relationship. Measurements were carried out at - 95 °C, under a *cw* laser-diode photoexcitation (405 nm, 25 µW), with an estimated photon flux absorbed by the perovskite $F \approx 10^{16} \text{cm}^{-2}\text{s}^{-1}$.

**Discussion**

To understand the observed modulation of PL with the gate voltage in perovskite PLTs, we consider a model which includes: (i) the source of photocarriers - photoexcitation, the two channels of their decay - (ii) a bimolecular (electron-hole) recombination, and (iii) the carrier trapping, and (iv) the presence of gate-induced mobile holes at the perovskite-dielectric interface of these transistors. **Figure 4a** schematically shows the processes considered in our model which is based on the following assumptions, in line with the current understanding of the optoelectronic properties of single-crystalline MHPs:[36]

First, at sufficiently high temperatures, 3D metal-halide perovskites behave as non-excitonic materials, where photoexcitation generates free (mobile) electrons and holes responsible for the photoconductivity and PL.[28, 45] The PL originates from a radiative bimolecular (electron-hole) recombination occurring with a rate proportional to the product of steady-state concentrations of electrons and holes. Non-excitonic nature of these materials leads, in particular, to the peculiar correlated power exponents in the photoexcitation-intensity dependence of the photoconductivity and PL of crystalline lead-halide perovskites, with the photoconductivity, $\sigma_{PC}$, exhibiting a crossover from the power exponent 1 to 1/2, and $I_{PL}$ exhibiting a crossover from the power exponent 2 to 3/2.[46]

Second, as in classic FETs, the density, $n_G$, of the gate-induced mobile holes in the accumulation channel at the perovskite/dielectric interface in our transistors is controlled by the gate voltage, $V_G$, according to the relationship (see, e.g., Refs.[30, 44, 47, 48]):

$$n_G = \begin{cases} -\frac{C_i}{e}(V_G - V_T), & V_G < V_T \\ 0, & V_G \geq V_T, \end{cases} \quad (1)$$

where $C_i$ is the gate-channel capacitance per unit area, $e$ is the elementary charge, $V_T$ is the threshold voltage.

Third, we assume for simplicity that photoexcitation generates equal numbers of electrons and holes, so that the photogenerated carrier density is simply denoted as $n$ ($n = n_e = n_p$). The photon-to-charge-carrier conversion factor (i.e., the photocarrier generation quantum efficiency) is denoted as $\kappa$ ($0 < \kappa \leq 1$).

Fourth, there are four length scales to consider: (1) the perovskite film's thickness, $d_{film} \sim 1$ μm; (2) the light penetration length, inversely proportional to the absorption coefficient, $\alpha$, and defining the characteristic Beer-Lampert depth scale of the photocarrier generation region, $\alpha^{-1} \sim 100$ nm;[49, 50] (3) the thickness of the field-effect accumulation channel, $d_{FE} \sim$ few nm, due to electrostatic screening;[51] and, finally, (4) the carrier diffusion length, $L_{diff}$, that is governed by the carrier lifetime and diffusivity. Prior studies have shown that $L_{diff}$ in single-crystalline metal-halide perovskites can be very long ($L_{diff} \gg 1$ μm),[29] longer than our typical film thickness and much longer than the light penetration length or the accumulation channel thickness ($L_{diff} > d_{film} \gg \alpha^{-1} \gg d_{FE}$). Combined with a relatively efficient charge transport in these single-crystalline materials, this allows us to assume that, for the purpose of *cw* experiments, photocarriers generated in the top portion of the film (within $\alpha^{-1}$ below the interface) quickly spread throughout the entire bulk (of thickness $d_{\text{film}}$), almost instantaneously establishing a homogeneous spatial distribution with an effective average density, $n$.

Fifth, unlike photocarriers in the bulk, the gate-induced holes are confined to the very thin accumulation layer formed at the interface with the gate dielectric ($d_{\text{FE}} \ll d_{\text{film}}$). Nevertheless, owing to the long carrier diffusion length and efficient charge transport, photogenerated electrons and holes can reach the film's surfaces, perhaps multiple times during their lifetime, and can thus interact with the gate-induced holes at the top interface of the device (via a surface electron-hole recombination) or be captured by interfacial traps. This allows us to consider the bulk photocarriers as effectively interacting with the interfacial gate-induced holes, such that these populations can be treated in the rate equation below as occupying the same physical volume.

Finally, the model does not consider ionic drifts and associated screening effects, which limits its applicability to situations where such effects are not dominant (e.g., at reduced temperatures, or in dense single-crystalline materials with reduced disorder).

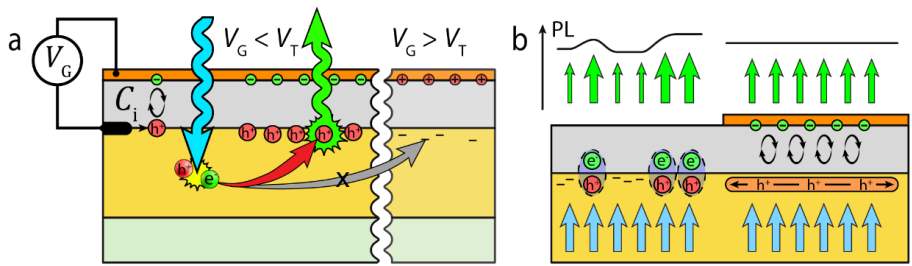

**Figure 4. Microscopic processes involved in the PL gating effect in perovskite transistors.** (**a**) The blue and green arrows represent the light incident on the transistor's active material through a semi-transparent top gate and the PL emitted by the material, respectively. The density of mobile holes (red circles) in the PLT's accumulation channel is controlled by the gate. Increasing the mobile hole density at the interface boosts the rate of bimolecular (electron-hole) recombination for the bulk photocarriers, leading to an increase in the PL intensity. (**b**) Unlike a spatially uniform PL emission for an idealized homogeneous distribution of delocalized charges at the interface (the right part of the cartoon), spatial inhomogeneities of trap distribution or interfacial doping at the realistic perovskite-dielectric interface (the left part) may lead to spatial fluctuations of the local threshold voltage and thus the local mobile carrier concentration in these transistors, leading, in turn, to a broadening of the bimolecular recombination rate and the resultant PL intensity over the channel area. This can be taken into account in our model via a Gaussian surface averaging (Eq. 6).

Based on these assumptions, the following rate equation for the projected (areal) steady-state photocarrier density, $n$, can be written:

$$\frac{dn}{dt} = \kappa F - \gamma(n_\text{G} + n)n - \tau^{-1}n = 0, \qquad (2)$$

where $F$ is the photon flux absorbed by the material, $\gamma$ is the coefficient of bimolecular (electron-hole) recombination, and $\tau$ is the trap-limited carrier lifetime. Note that for convenience we wrote this equation for the projected (areal) carrier density, so that both $n$ and $n_\text{G}$ are in [cm$^{-2}$], $F$ is the photon flux in [cm$^{-2}$s$^{-1}$], and $\gamma$ is in [cm$^2$s$^{-1}$]. Thus defined parameters are equivalent to their conventional (3D) analogs normalized by the effective film thickness, which is justified by the above relationships between the relevant length scales ($L_\text{diff} > d_\text{film} \gg \alpha^{-1} \gg d_\text{FE}$). Equation 2 features the photocarrier source (photoexcitation at a rate $\kappa F$) and the two channels of carrier decay – a bimolecular (electron-hole) recombination and trapping. Under steady-state conditions, the steady-state concentration of mobile photocarriers, $n$, in Equation 2 is constant ($\mathrm{d}n/\mathrm{d}t = 0$), and the Equation has the following analytical solution:

$$n(F, n_\mathrm{G}) = \sqrt{\left(\frac{1+\gamma\tau n_\mathrm{G}}{2\gamma\tau}\right)^2 + \frac{\kappa}{\gamma}F} - \frac{1+\gamma\tau n_\mathrm{G}}{2\gamma\tau}. \tag{3}$$

At relatively high temperatures, the PL in 3D metal-halide perovskites mainly originates from a radiative recombination of mobile electrons and holes.[45, 46, 52] Thus, $I_\mathrm{PL}$ is proportional to the bimolecular recombination decay term in Equation 2:

$$I_\mathrm{PL} \propto \gamma(n_\mathrm{G} + n)n, \tag{4a}$$

Which Eq. 2 allows to rewrite as:

$$I_\mathrm{PL} \propto \kappa F - \tau^{-1}n. \tag{4b}$$

We can thus express the steady-state PL intensity, $I_\mathrm{PL}$, via the photoexcitation flux density, $F$, and the $V_\mathrm{G}$-dependent photocarrier density, $n$, with the help of Equations (1), (3), and (4). In addition, for compactness, the resultant $I_\mathrm{PL}$ can be normalized by its maximum value ($I_\mathrm{PL}^\mathrm{max} = \kappa F$), yielding the final expression for the normalized PL intensity as a function of $V_\mathrm{G}$:

$$\frac{I_\mathrm{PL}(V_\mathrm{G})}{I_\mathrm{PL}^\mathrm{max}} = \begin{cases} 1 - \sqrt{\left(\frac{1-\gamma\tau C_\mathrm{i}(V_\mathrm{G}-V_\mathrm{T})/e}{2\gamma\tau^2\kappa F}\right)^2 + \frac{1}{\gamma\tau^2\kappa F}} + \frac{1-\gamma\tau C_\mathrm{i}(V_\mathrm{G}-V_\mathrm{T})/e}{2\gamma\tau^2\kappa F}, & V_\mathrm{G} < V_\mathrm{T} \\ 1 - \sqrt{\left(\frac{1}{2\gamma\tau^2\kappa F}\right)^2 + \frac{1}{\gamma\tau^2\kappa F}} + \frac{1}{2\gamma\tau^2\kappa F}, & V_\mathrm{G} \geq V_\mathrm{T}. \end{cases} \tag{5}$$

This analytical solution can be used directly to fit the experimental data such as those shown in Fig. 3b (an example fit is shown in Supplementary sec. 5). The model correctly captures the main experimental observations, including the increase in $I_\mathrm{PL}$ with decreasing $V_\mathrm{G}$, the saturation of the effect at sufficiently high negative $V_G$, and $I_{PL}$ leveling off at $V_G > V_T$, in the transistor's off state.

To improve the model fit in the subthreshold region, we recall that, unlike electrical FETs, where a single observable representing the entire accumulation channel is the source-drain current parametrized via a specific (fixed) threshold voltage, $V_T$, the photoluminescence in PLTs is collected from a macroscopic channel area of the device, and its intensity, $I_{PL}$, is an integral quantity. The PL microscopy images and videos (Figs. 1b, S5, and Supplementary videos) show that the PL is not perfectly homogeneous over the channel area even in these high-quality single crystals. Spatial fluctuations in $I_{PL}$ might originate from (a) variations in the local density of traps, or (b) inhomogeneous charge-transfer doping at the perovskite/parylene-N interface. To account for such fluctuations, we propose that the corresponding *local* threshold voltage, $V_T$, in the channel is distributed around a mean, $\overline{V}_T$, following a Gaussian distribution of width (i.e., standard deviation) $\Delta V_T$. We can thus incorporate the following Gaussian averaging of the $I_{PL}(V_G, V_T)$ function given by Equation 5 over the channel area into the fitting procedure of our experimental data:

$$I_{PL}^{fit}(V_G)_{\overline{V}_T, \Delta V_T} = \frac{1}{\Delta V_T \sqrt{2\pi}} \int I_{PL}(V_T) e^{-(V_T - \overline{V}_T)^2 / 2(\Delta V_T)^2} \, dV_T. \tag{6}$$

**Figure 3b** shows the corresponding fit (black solid line) of the experimental data (symbols). Both hysteresis directions are fitted simultaneously via a global fit, revealing a very good overall consistency. The corresponding Origin C code used for this fitting is included as a Supplementary file. The mean (averaged over the channel area) threshold voltage and the width of its distribution obtained from the fit are $\overline{V}_T = 27 \pm 3$ V and $\Delta V_T = 10 \pm 2$ V. The average density of mobile holes due to the interfacial charge-transfer doping at the perovskite/parylene-N interface in these devices can thus be estimated as ~ $3.0\times10^{11}$ cm$^{-2}$. The details of possible inhomogeneities in the distributions of trap density and/or doping over the channel area in epitaxial CsPbBr$_3$ PLTs might be a subject of further studies.

Fitting the experimental data with this model allows rough estimates of the microscopic parameters of photocarrier dynamics, including the bimolecular recombination coefficient, $\gamma$ = 1.5x10$^{-4}$ cm$^2$/s, and the trap-limited carrier lifetime, $\tau$ = 1x10$^{-6}$ s. Additionally, normalizing the PL intensity by its theoretical maximum ($I_\mathrm{PL}^\mathrm{max} = \kappa G$) gives the relative PL yield originating from the bimolecular recombination: the left vertical axis of Fig. 3b shows this yield in percent. Given that photoexcitation of CsPbBr$_3$ at these relatively high temperatures is still completely non-excitonic,[28] and that the thickness of the active material in our PLTs greatly exceeds the incident light penetration length ($d_\mathrm{film} \gg \alpha^{-1}$), ensuring near-complete absorption of incident photons, we can safely assume that 100% on this axis corresponds to a single PL photon emitted per incident excitation photon. This consideration shows that at sufficiently high negative $V_\mathrm{G}$, we can reach the *external* PL quantum yield in our PLTs approaching 100%. We emphasize that reaching a PLQY of nearly 100% in a solid-state material, simultaneously with it exhibiting an intrinsic (not dominated by disorder) charge transport in the dark (as demonstrated in CsPbBr$_3$ FETs), is a remarkable feast of material growth optimization and interface engineering.

It would be useful to consider the two limiting cases of the above model in terms of trapping rates and discuss carrier lifetimes. When trapping is negligible ($\tau \to \infty$), the rate Equation 2 leads to a PL intensity proportional to the photoexcitation density $F$ but independent of the gate voltage $V_\mathrm{G}$, while in the opposite limit of a significant trapping ($\tau \to 0$), the first-order

Taylor expansion of Equation 5 shows that $I_{PL}$ is proportional to the product of $F$ and $V_G$:

$$I_{PL} \to \begin{cases} \kappa F, & \tau \to \infty \text{ (negligible trapping)}, \\ \tau\gamma \cdot \kappa F \cdot \frac{C_i(V_G - V_T)}{e}, & \tau \to 0 \text{ (non-negligible trapping)}. \end{cases} \quad (7)$$

These limiting cases make sense for the PL originating from a radiative electron-hole recombination. Indeed, when carrier trapping is negligible, *all* photogenerated electrons and holes remain mobile until they eventually interact with each other and recombine, making additional gate-induced holes unnecessary for boosting the PL. On the contrary, when carrier trapping is non-negligible, only a fraction of the photocarriers remains mobile and available for recombination. Under such conditions, adding extra mobile holes by gating in a FET geometry positively contributes to the net electron-hole recombination rate and boosts the PL.

Finally, because the PL intensity is given by $I_{PL} \propto \gamma(n_G + n)n$ (Eq. 4), at first glance, it may appear that $I_{PL}$ should *always* be increasing with an increasing population, $n_G$, of the gate-induced mobile holes, even when trapping is negligible ($\tau \to \infty$) and the photocarrier lifetime is dominated by the bimolecular (electron-hole) recombination. Such a perceived trend, however, clashes with the first limiting case in Equation 7 (at $\tau \to \infty$, $I_{PL}$ should be independent of $n_G$ and, thus, $V_G$). This apparent contradiction is resolved by recalling that, under steady-state conditions, a dynamic equilibrium between the photogeneration rate and the rate of bimolecular recombination sets in, which governs the steady-state photocarrier population, $n$: adding more holes ($n_G$) to the system transiently increases the bimolecular recombination rate, which, in turn, decreases the steady-state photocarrier population, $n$. Thus, in the limit $\tau \to \infty$, the self-consistent steady-state solution is a fixed ($V_G$-independent) PL intensity, $I_{PL} \to \kappa F$, corresponding to the saturation of the gating effect observed at high negative $V_G$ (Fig. 3b). In such a regime, adding more holes by gating ($n_\text{G}$) leads to a reduction in the steady-state photocarrier concentration $n$, while leaving the product $\gamma(n_\text{G} + n)n$ unchanged.

These considerations show that some residual trapping ($\tau^{-1}n$ term in Eq. 2) is necessary for the observation of the PL gating effect. With this term present (i.e., $\tau$ being finite), adding more holes by gating ($n_\text{G}$) not only reduces the steady-state photocarrier population, $n$, but also suppresses the corresponding non-radiative decay rate, $\tau^{-1}n$. This ensures a higher steady-state photocarrier population, $n$, compared to the case of negligible trapping, thus leading to the term $\gamma(n_\text{G} + n)n$ increasing with $n_\text{G}$, rather than remaining constant. Simultaneously, the bimolecular-recombination dominated carrier lifetime, $\tau_\text{bimol}$, would be decreasing with $n_\text{G}$ in this regime. $\tau_\text{bimol}$ is inversely proportional to the average carrier concentration and can be defined as:

$$\tau_\text{bimol} \equiv \frac{1}{\gamma \cdot \sqrt{(n_\text{G}+n)n}}. \tag{8}$$

When the $\tau^{-1}n$ term in the rate Equation 2 is non-negligible ($\tau$ is finite), applying proper gate voltage ($V_\text{G} < V_\text{T}$) increases the mobile hole concentration by $n_\text{G}$ and skews the competition between the two channels of carrier decay from the non-radiative trapping towards the radiative bimolecular recombination. Thus, gating a PLT in this regime affects the effective photocarrier lifetime: with increasing $n_\text{G}$, $\tau_\text{bimol}$ given by Equation 8 becomes progressively smaller, until it falls below the fixed trap-limited lifetime, $\tau$, and starts defining the overall effective carrier lifetime. In such a case, more photocarriers would radiatively recombine before being trapped, leading to an enhancement in PL with $n_\text{G}$. When $\tau_\text{bimol}$ becomes significantly smaller than $\tau$, which essentially corresponds to the limit $\tau \to \infty$, a saturation of the PL gating effect should occur, which we experimentally observe at sufficiently high negative $V_G$ (Fig. 3b). Such a saturation is more readily observed at low temperatures, because the increased carrier mobility boosts the bimolecular recombination rate, helping to reach $\tau_{bimol} \ll \tau$. These tendencies of the model are plotted in Supplementary sec. 6; they reveal that the relative magnitude of the PL gating effect, as well as the effect of $V_G$ on $\tau_{bimol}$, progressively diminish with increasing $\tau$ (that is, with the system approaching the perfectly trap-free limit, $\tau \to \infty$).

**Summary**

In conclusion, we have demonstrated an electrostatic photoluminescence gating effect in all-solid-state $CsPbBr_3$ perovskite field-effect transistors. Under *cw* photoexcitation in the visible range, PL of the active material in these devices can be reversibly modulated with a gate voltage by nearly 100%. The microscopic mechanism of the effect appears to stem from a contribution of the gate-induced mobile holes to the bimolecular (electron-hole) recombination of photocarriers, thus affecting the average steady-state photocarrier population and lifetime. High-quality epitaxial single-crystalline $CsPbBr_3$, used as the semiconductor in these devices, enabled the efficient charge transport and high photoluminescence efficiency, which is important for the operation of these *photoluminescence transistors*. We propose an analytical model of the effect that captures the main experimental observations and allows estimating the key microscopic parameters of photocarrier dynamics. Reversibly and purely electrostatically (that is, without drawing current) tuning the photoluminescence of a semiconductor with an "electric knob" provides new exciting opportunities for the fundamental and applied research in optoelectronics of emerging materials. It can potentially contribute to applications in optical switching, sensing, lasing, as well as optical integrated circuits, and telecommunications.

**Methods**

Details of CsPbBr3 epitaxial crystal growth can be found in Supplementary Section 1.1. In brief, cesium bromide (CsBr) and lead(II) bromide ($PbBr_2$) precursors were used to perform the growth of stoichiometric crystalline perovskite film on mica carried out in a custom-designed vapor-phase epitaxial growth system in helium gas, at a flow rate of 100 sccm, pressure of 0.1 bar, and in the temperature range 540 - 500 °C. The thickness of the obtained large-area crystalline films was in the range 0.4 – 1.2 µm, ensuring that nearly 100% of the incident light in the visible range is absorbed by the perovskite (optical density > 2).

Fabrication of the photoluminescence transistors parallels the procedure that we have reported for the epitaxial single-crystal $CsPbBr_3$ FETs,[30] except for the optically semi-transparent gate electrode used here. In brief, epitaxial $CsPbBr_3$ films were hand-patterned to macroscopic (~ few mm) single-crystalline rectangles; colloidal graphite was used to paint contacts; parylene-N of thickness in the range 1 - 1.3 µm was used as a gate dielectric;[40, 53] ultra-thin (3 - 10 nm-thick), semi-transparent, yet conductive gold film was thermally evaporated through a shadow mask to define the top gate electrode (for details, see Supplementary sec. 1.2 and 7).

For the gated PL measurements, devices were placed in a cryogenic optical vacuum chamber. $V_G$ in the range $-50 \div 50$ V was applied using Keithley 2400 or 2450 Source-Meters relatively to the grounded contacts. These measurements were carried out in *two* physically different and dissimilar setups, one located at Rutgers (USA) and another one at Imperial College London (UK), operating at different temperatures: - 20 ÷ 20 °C at Rutgers, and - 95 °C at the Imperial (Fig. S2). The details of the experimental setups are given in Supplementary sec. 1.3.

**Acknowledgements**

VP and VB acknowledge funding from the U.S. Department of Energy, Office of Basic Energy Sciences, Division of Materials Sciences and Engineering under Award DE-SC0025401 (material synthesis, device fabrication, and measurements), and the Donald H. Jacobs Chair in Applied Physics (equipment upgrades).  AAB and BH acknowledge funding from UKRI/EPSRC (ActionSpec, grant ref: EP/X030822/1).